\documentclass[a4paper,fleqn,usenatbib]{mnras}

\usepackage{newtxtext,newtxmath}
\usepackage[flushleft]{threeparttable}
\usepackage{times,epsfig,natbib,amssymb,amsmath,graphics,longtable,lscape,threeparttable}

\usepackage[T1]{fontenc}
\usepackage{ae,aecompl}

\usepackage{graphicx}	% Including figure files
\usepackage{amsmath}	% Advanced maths commands
\usepackage{nccmath}
\usepackage{amssymb}	% Extra maths symbols
\usepackage{pdflscape}	% table landscape
\usepackage{multicol}
\title[$H_0$ from SNe~II]{A measurement of the Hubble constant from Type II supernovae}

\author[de Jaeger et al.]
{T. de Jaeger$^{1,2}$\thanks{E-mail: tdejaeger@berkeley.edu},
B. E. Stahl$^{1,3,4}$,
W. Zheng$^{1}$,
A. V. Filippenko$^{1,5}$, 
A. G. Riess$^{6,7}$,
\newauthor
L.~Galbany$^{8}$
\\
\small
$^{1}$Department of Astronomy, University of California, Berkeley, CA 94720-3411, USA.\\
$^{2}$Bengier Postdoctoral Fellow.\\
$^{3}$Department of Physics, University of California, Berkeley, CA 94720-7300, USA.\\
$^{4}$Marc J. Staley Graduate Fellow.\\
$^{5}$Miller Senior Fellow, Miller Institute for Basic Research in Science, University of California, Berkeley, CA 94720, USA.\\
$^{6}$Space Telescope Science Institute, 3700 San Martin Drive, Baltimore, MD 21218, USA.\\
$^{7}$Department of Physics \& Astronomy, Johns Hopkins University, Baltimore, MD 21218, USA.\\
$^{8}$Departamento de F\'isica Te\'orica y del Cosmos, Universidad de Granada, E-18071 Granada, Spain.\\
}

\date{}

\pubyear{2020}

\begin{document}
\label{firstpage}
\pagerange{\pageref{firstpage}--\pageref{lastpage}}

\maketitle

% Abstract of the paper (Aims, Methods, and Results)
\begin{abstract}
\noindent
Progressive increases in the precision of the Hubble-constant measurement via Cepheid-calibrated Type Ia supernovae (SNe~Ia) have shown a discrepancy of $\sim 4.4\sigma$ with the current value inferred from {\it Planck} satellite measurements of the cosmic microwave background radiation and the standard $\Lambda$CDM cosmological model. This disagreement does not appear to be due to known systematic errors and may therefore be hinting at new fundamental physics.
Although all of the current techniques have their own merits, further improvement in constraining the Hubble constant requires the development of as many independent methods as possible. In this work, we use SNe~II as standardisable candles to obtain an independent measurement of the Hubble constant. Using 7 SNe~II with host-galaxy distances measured from Cepheid variables or the tip of the red giant branch, we derive H$_0= 75.8^{+5.2}_{-4.9}$\,km\,s$^{-1}$\,Mpc$^{-1}$ (statistical errors only). Our value favours that obtained from the conventional distance ladder (Cepheids + SNe~Ia) and exhibits a difference of 8.4\,km\,s$^{-1}$\,Mpc$^{-1}$ from the {\it Planck} $+\Lambda$CDM value. Adding an estimate of the systematic errors (2.8\,km\,s$^{-1}$\,Mpc$^{-1}$) changes the $\sim 1.7\sigma$ discrepancy with {\it Planck} $+\Lambda$CDM to $\sim 1.4\sigma$. Including the systematic errors and performing a bootstrap simulation, we confirm that the local H$_0$ value exceeds the value from the early Universe with a confidence level of 95\%. As in this work we only exchange SNe~II for SNe~Ia to measure extragalactic distances, we demonstrate that there is no evidence that SNe~Ia are the source of the H$_0$ tension.

\end{abstract}


%%%%%%%%%%%%%%%%%%%%%%%%%%%%%%%%%%%%%%%%%%%%%%%%%%

%%%%%%%%%%%%%%%%% BODY OF PAPER %%%%%%%%%%%%%%%%%%
\section{Introduction}

The current expansion rate of the Universe, known as the Hubble constant (H$_0$), remains one of the most important parameters in modern cosmology. Determining an accurate value is essential for obtaining information regarding our Universe, including its age and evolution. Since the discovery of the expansion of the Universe \citep{lemaitre27,hubble29}, many efforts have been made to measure H$_0$ precisely and decrease its uncertainties. 
Traditionally, there have been two main routes for determining H$_0$. First, H$_0$ can be measured locally through the distance-ladder method. Distances to galaxies in the Hubble flow, where peculiar velocities are insignificant (the motion of galaxies is almost entirely due to the expansion of the Universe), can be measured using Type Ia supernovae (SNe~Ia; e.g., \citealt{min41,elias85,filippenko97,howell11}, and references therein). 
For sufficiently nearby SN~Ia host galaxies where individual stars can be resolved with the {\it Hubble Space Telescope (HST)}, the distances can be determined (and hence the peak luminosities of the SNe~Ia can be calibrated) through measurements of Cepheid variable stars \citep{freedman01,sandage06,riess09,freedman10,riess11,riess16,riess18a,riess18b,burns18,dhawan18,riess19} or the tip of the red giant branch (TRGB) in the Hertzsprung-Russell diagram \citep{madore09,jang17a,jang17b,freedman19,yuan2019}.
Cepheids and TRGBs can, in turn, be calibrated to geometric anchor distances like Milky Way Cepheid parallaxes \citep{benedict07,riess14,casertano16}, Keplerian motion of masers in NGC 4258 \citep{humphreys13,reid19}, or detached eclipsing binary stars in the Large Magellanic Cloud \citep{pietrzynski13}. Using this distance-ladder technique, the uncertainty in the measurement of H$_0$ has improved from $\sim 10$\% \citep{freedman01} to $<2$\% \citep{riess19} during the last 20\,yr. Currently, the most precise estimate of the H$_0$ is $74.03 \pm 1.42$\,km\,s$^{-1}$\,Mpc$^{-1}$ \citep{riess19}.

Second, H$_0$ can be predicted using the sound horizon observed from the cosmic microwave background radiation (CMB; e.g., \citealt{fixsen96,jaffe01,spergel07,bennett03,planck18}). However, H$_0$ cannot be constrained directly from CMB observations. While the local H$_0$ value at redshift $z \approx 0$ is obtained using the distance ladder from Hubble-flow SNe~Ia ($z \approx 0.02$--0.15) using calibrated objects for each step, the second value is obtained using data at $z \approx 1100$ and extrapolated to $z \approx 0$ based on the physics of the early Universe. Assuming a flat $\Lambda$ cold dark matter ($\Lambda$CDM) cosmological model, \citet{planck18} obtained a value of H$_0 = 67.4 \pm 0.5$\,km\,s$^{-1}$\,Mpc$^{-1}$. A consistent value is also found when an intermediate-redshift rung and the inverse distance ladder method are used. For exemple, using baryon acoustic oscillations (BAO) calibrated from the CMB to anchor SNe~Ia at $z > 0.1$, \citet{macaulay19} obtained H$_0 = 67.8 \pm 1.3$\,km\,s$^{-1}$\,Mpc$^{-1}$.

Although the two H$_0$ values from opposite ends of the Universe agree to within 10\%, their error bars do not overlap and a discrepancy of $\sim 4.4\sigma$ is seen with SNe~Ia \citep{riess19} (and even greater significance, exceeding $6\sigma$, when other nearby-universe techniques are combined; \citealt{riess20}). This disagreement does not appear to be due to known systematic errors. Independent reanalyses of the \citet{riess16} data have shown minimal differences \citep{cardona17,zhang17,feeney18,follin2018,dhawan18,burns18}, and as more data and independent methods are used the discrepancy is increasing \citep{riess19}. This significant tension between both measurements could arise from relativistic particles (dark radiation), nonzero curvature, early dark energy, increasing dark energy, or new fundamental physics.

However, to confirm the ``H$_0$ tension'' it is important to develop as many independent methods as possible having different systematic errors (see \citealt{riess20} for a recent review). For example, a novel measurement of H$_0$ has been made using quasars that are strongly gravitationally lensed into multiple images and the time-delay distance technique \citep{bonvin17,wong2019}. From their analysis of multiply-imaged quasars, the H0LiCOW collaboration \citep{bonvin17} has measured a value of $73.3^{+1.7}_{-1.8}$\,km\,s$^{-1}$\,Mpc$^{-1}$ \citep{wong2019}, consistent with the most recent local distance-ladder measurements \citep{riess19}. Additionally, the Megamaser Cosmology Project, using geometric distance measurements to megamaser-hosting galaxies, obtained an independent H$_0$ value of $73.9 \pm 3.0$\,km\,s$^{-1}$\,Mpc$^{-1}$ \citep{pesce2020}. Other techniques such as ``standard sirens'' from merger events detected through gravitational waves \citep{abbott17} are also promising, but their current precision is not sufficient to put strong constraints; there is only a single electromagnetic gravitational-wave counterpart detection (H$_0 = 70^{+12}_{-8}$\,km\,s$^{-1}$\,Mpc$^{-1}$). However, even ``dark sirens'' that have no known electromagnetic detection have potential \citep{vasyleyv20}.

Needing approaches independent of  SNe~Ia, we study the use of SNe~II as cosmological standardisable candles \citep{dejaeger20}. Observationally, SNe~II are characterised by the presence of strong hydrogen features in their spectra (see \citealt{filippenko97,filippenko00} and \citealt{galyam17} for overviews), and a plateau of varying steepness in their light curves \citep{barbon79,anderson14a,galbany16a} during the hydrogen recombination phase. Their use as cosmic distance indicators is mainly motivated by the fact that they are more abundant than SNe~Ia \citep{li2011,graur17b} and their progenitors and environments are better understood than those of SNe~Ia. It is now accepted that SN~II progenitors arise from only one stellar population (red supergiant stars) for which the explosion mechanism is reasonably well understood \citep{woosley95,janka01,janka07}.

At first sight, the SN~II family displays a large range of peak luminosities (more than 2\,mag); however, as for SNe~Ia, their luminosities can be calibrated. To date, different theoretical and empirical SN~II distance-measurement methods have been proposed and tested \citep[e.g.,][and references therein]{nugent17}. First, the expanding photosphere method based on the relation between the angular size and the ratio between its observed and theoretical flux was developed by \citet{kirshner74}. Following this method, several empirical methods have developed: the standard candle method based on the correlation between the luminosity and the expansion velocities (SCM; \citealt{hamuy02,dejaeger20}), the photospheric magnitude method which corresponds to a generalisation of the SCM (PMM; \citealt{rodriguez14,rodriguez19a}), and the photometric colour method which uses the relation between the luminosity and the slope of the plateau (PCM; \citealt{dejaeger15b,dejaeger17a}). Using those techniques, H$_0$ values of $73 \pm 13$\,km\,s$^{-1}$\,Mpc$^{-1}$ (EPM; \citealt{schmidt94}), $69 \pm 16$\,km\,s$^{-1}$\,Mpc$^{-1}$ (SCM; \citealt{olivares10}), and $\sim 71 \pm 8$\,km\,s$^{-1}$\,Mpc$^{-1}$ (PMM in the $V$ band; \citealt{rodriguez19a}) have been derived, but their precisions are not yet comparable to those of {\it Planck} or SNe~Ia owing to a lack of SNe~II in the Hubble flow, as well as to a small number of Cepheids or resolved red giants in SN~II host galaxies. 

In this work, we increase the number of calibrators and use the largest SN~II sample in the Hubble flow to derive H$_0$ with a precision of $\sim 6.5$\% (statistical). Section 2 contains a description of the SN data, and in Section 3 we present the methods used to derive H$_0$. We discuss our results in Section 4 and summarise our conclusions in Section 5.

\section{Data sample}

For our analysis, we use SNe~II from different surveys: the Carnegie Supernova Project-I (CSP-I\footnote{\url{http://csp.obs.carnegiescience.edu/}}; \citealt{ham06}), the Lick Observatory Supernova Search (LOSS) with the 0.76\,m Katzman Automatic Imaging Telescope (KAIT\footnote{\url{http://w.astro.berkeley.edu/bait/kait.html}}; \citealt{filippenko01}), the Sloan Digital Sky Survey-II SN Survey (SDSS-II\footnote{\url{http://classic.sdss.org/supernova/aboutsupernova.html}}; \citealt{frieman08}), the Supernova Legacy Survey (SNLS\footnote{\url{http://cfht.hawaii.edu/SNLS/}}; \citealt{astier06,perrett10}), the Dark Energy Survey Supernova Program (DES-SN\footnote{\url{https://portal.nersc.gov/des-sn/}}; \citealt{bernstein12}), and the Subaru Hyper-Suprime Cam Survey (SSP-HSC; \citealt{miyazaki12,aihara18a}). 
We also add SN~2009ib \citep{takats15}, a nearby SN~II for which we have a SN host distance measurement from Cepheids estimated by the SH0ES\footnote{``Supernovae, H$_0$ for the Equation of State of Dark Energy''; \citet{riess11}} team (A. G. Riess, 2020, private communication). All of the data have already been used in different cosmological studies \citep{poznanski09,poznanski10,andrea10,dejaeger15b,dejaeger17a,dejaeger17b,dejaeger20} and a complete description of the surveys is given by \citet{poznanski09} (KAIT-P09), \citet{andrea10} (SDSS-SN), \citet{dejaeger15b} (CSP-I), \citet{dejaeger17b} (SNLS, HSC), \citet{ganeshalingam10,stahl19a,dejaeger19} (KAIT-d19), and \citet{dejaeger20} (DES-SN).
 
Following \citet{dejaeger20}, our method is applied at 43\,d after the explosion, and only SNe~II with an explosion date uncertainty smaller than 10\,d are selected. Additionally, since in this work we use the SCM, at least one spectrum per SN is needed to measure the photospheric expansion velocity. 
After these cuts, our sample consists of 125 SNe~II: 49 (CSP-I) $+$ 13 (SDSS-SN) $+$ 4 (SNLS) $+$ 12 (KAIT-P09) $+$ 30 (KAIT-d19) $+$ 15 (DES-SN) $+$ 1 (HSC) $+$ 1 (SN~2009ib). Among these 125 SNe~II, 89 SNe~II have $z > 0.01$.

Note that all of the CMB redshifts ($z_{\rm CMB}$) were obtained from the NASA/IPAC Extragalactic Database (NED\footnote{\url{http://ned.ipac.caltech.edu/}}) and were corrected to account for peculiar flows ($z_{\rm corr}$) induced by visible structures using the model of \citet{carrick15}. A residual peculiar velocity uncertainty of 250\,km\,s$^{-1}$ is also assumed.

\section{Methods}

\subsection{SN~II standardisation}

To calibrate SNe~II, the SCM is applied; see \citet{dejaeger20} for a full description of the method. Briefly, we use the observed correlations between SN~II luminosity and photospheric expansion velocity during the plateau phase as well as the colour to correct the SN~II magnitude. Thus, for each SN the corrected magnitude can be written as

\begin{ceqn}
\begin{align}
m^{X}_{\rm corr}=\mathrm{m}_{X}+\alpha\, \mathrm{\log_{10}}\left( \frac{v_\mathrm{H\beta}}{<v_\mathrm{H\beta}>} \right) - \beta (c - <c>)~,
\label{m_model}
\end{align}
\end{ceqn}
where $X$ is the $X$-band filter, $m$ is the apparent magnitude at 43\,d, $c$ is the colour ($<c>$ is the average colour), and $v_\mathrm{H\beta}$ is the velocity measured using H$\beta$ absorption ($<v_\mathrm{H\beta}>$ is the average value). To determine the best-fitting parameters ($\alpha$ and $\beta$) and to derive the Hubble diagram, a Monte Carlo Markov Chain (MCMC) simulation is performed using the Python package \textsc{EMCEE} developed by \citet{foreman13}. For more details, the reader is referred to Equations (1), (2), and (3) of \citet{dejaeger20}.

Note that all magnitudes were simultaneously corrected for Milky Way extinction ($A_{V,G}$; \citealt{schlafly11}), redshifts due to the expansion of the Universe (K-correction; \citealt{oke68,hamuy93,kim96,nugent02}), and differences between the photometric systems (S-correction; \citealt{stritzinger02}) using the cross-filter K-corrections defined by \citet{kim96}. More details regarding these correction are given by \citet{nugent02}, \citet{hsiao07}, \citet{dejaeger17a}, and references therein.

\subsection{Calibrators}\label{txt:calibrator}

In our low-redshift samples (CSP-I, KAIT, and SN~2009ib), six SNe~II have direct or indirect host-galaxy distance measurements from Cepheids, as follows. 

$\bullet$ SN~1999em in NGC~1637 for which the distance modulus derived by \citet{leonard03} has been updated using the new Large Magellanic Cloud (LMC) distance \citep{pietrzynski19} and an LMC to Milky Way Cepheid abundance difference of 0.30\,dex \citep{riess19}. The distance modulus used in this work is $\mu = 30.26 \pm 0.09$\,mag. Note that the metallicity term for optical Cepheid measurements used by \citet{leonard03} was 0.24\,mag\,dex$^{-1}$.

$\bullet$ SN~1999gi in NGC~3184 ($\mu=30.64 \pm 0.11$\,mag; updated from \citealt{leonard02b} using the new LMC distance), which is estimated through the average of the Cepheid distances of two galaxies (NGC~3319, $\mu = 30.60 \pm 0.08$\,mag; NGC~3198, $\mu = 30.68 \pm 0.08$\,mag; updated from \citealt{freedman01}) associated with the same galaxy group \citep{tully88}.

$\bullet$ SN~2005ay in NGC~3998 ($\mu =31.74 \pm 0.07$\,mag; \citealt{riess16}), which is measured indirectly from the Cepheid distance of NGC~3982, also a member of the Ursa Major Group.

$\bullet$ SN~2008bk ($\mu = 27.66 \pm 0.11$\,mag; \citealt{zgirski17}) in NGC~7793.

$\bullet$ SN~2009ib in NGC~1559 ($\mu =31.416 \pm 0.049$\,mag; A. G. Riess, 2020, private communication). This value is consistent with the one derived by \citet{huang20} using Mira variable stars ($\mu =31.41 \pm 0.05$ (stat) $\pm~0.052$ (sys)\,mag

$\bullet$ SN~2012aw in NGC~3351 ($\mu = 29.82 \pm 0.09$\,mag;) updated from \citet{kanbur03} following the prescriptions used for SN~1999em in NGC~1637.\\

%\noindent
In addition to these six SNe~II with Cepheid measurements, we also have three SNe~II with TRGB distance derivations.

$\bullet$ SN~2004et in NGC~6946 ($\mu = 29.38 \pm 0.09$\,mag; \citealt{anand18}).

$\bullet$ SN~2005cs in NGC~5194 (M51; $\mu = 29.62 \pm 0.09$\,mag; updated from \citealt{mcquinn17}).

$\bullet$SN~2013ej in NGC~628 (M74; $\mu = 29.90 \pm 0.10$\,mag; updated from \citealt{mcquinn17}).

For SN~2005cs and SN~2013ej, to convert the TRGB luminosities to distance moduli, we used a zero-point calibration of $-4.01$ from outer and halo fields in NGC 4258 \citep{reid19} instead of $-4.06$ for typical TRGB colour \citep{rizzi07}.

A summary of all the calibrators available in this work can be found in Table \ref{tab:calibrators}. To homogenise our calibrators, we remove two objects from this analysis. First, SN~2008bk, a low-luminosity SN~II \citep{vandyk12}, the only object for which the distance was obtained using ground-based (and not {\it HST}) observations (and then only of 11 Cepheids), and we cannot determine if these zero-points are consistent with those from {\it HST}. Second, SN~2004et has the largest Milky Way extinction ($A_{V,G} \approx 1.0$\,mag), making it very unreliable. Note that \citet{rodriguez19a} also identified SN~2004et as an outlier and they removed it from their calibrator sample.

\begin{table*}
\center
\caption{Calibrator sample.} 
\begin{tabular}{lccccc}
\hline
SN name	& Host Galaxy & $\mu$ (mag) & calibrator &Used &references  \\
\hline
SN~1999em &NGC~1637 &30.26 $\pm$ 0.09 & Cepheids & Yes &Updated from \citet{leonard03}\\
SN~1999gi &NGC~3184 &30.64 $\pm$ 0.11 & Cepheids & Yes &Updated from \citet{leonard02b}\\
SN~2004et &NGC~6946 &29.38 $\pm$ 0.09 & TRGB & No  &\citet{anand18}\\
SN~2005ay &NGC~3998 &31.74 $\pm$ 0.07 & Cepheids & Yes  &\citet{riess16}\\
SN~2005cs &NGC~5194/M51 &29.62 $\pm$ 0.09 & TRGB & Yes  & Updated from \citet{mcquinn17}\\
SN~2008bk &NGC~7793 &27.66 $\pm$ 0.11 & Cepheids & No  &\citet{zgirski17}\\
SN~2009ib &NGC~1559 &31.42 $\pm$ 0.05 & Cepheids & Yes  &A. G. Riess (2020), priv. comm.\\
SN~2012aw &NGC~3351 &29.82 $\pm$ 0.09 & Cepheids & Yes  &Updated from \citet{kanbur03}\\
SN~2013ej &NGC~628/M74  &29.90 $\pm$ 0.10 & TRGB & Yes  &Updated from \citet{mcquinn17}\\
\hline
\label{tab:calibrators}
\end{tabular}
\end{table*}

\subsection{H$_0$ derivation from SNe~II}\label{txt:H0_SNII}

\begin{figure}
	\includegraphics[width=1.0\columnwidth]{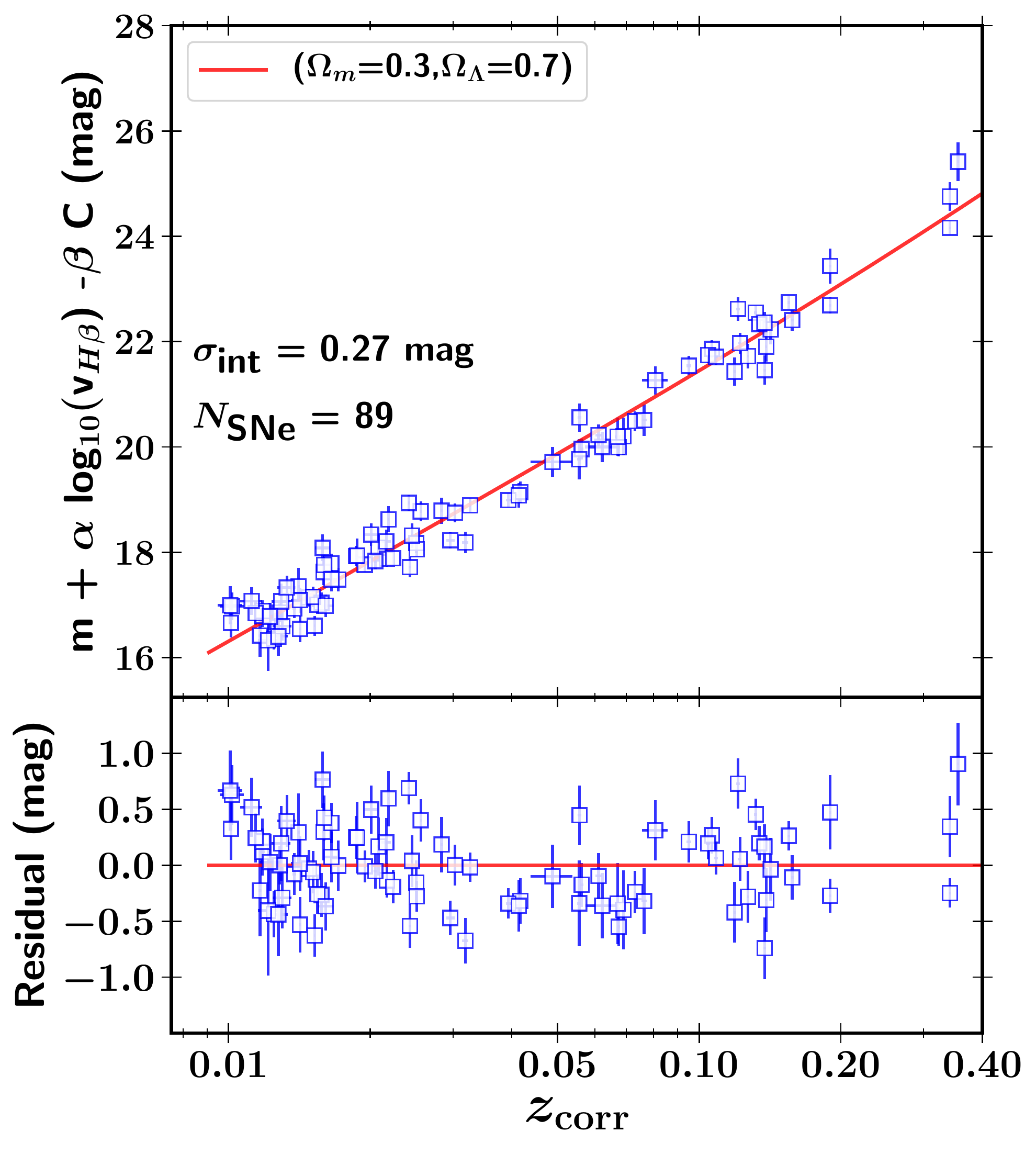}
\caption{Hubble diagram (top) and residuals from the $\Lambda$CDM model (bottom) using the SCM as applied to our sample of 89 SNe~II in the Hubble flow. $z_{\rm corr}$ corresponds to the CMB redshifts corrected to account for peculiar flows. The red solid line is the Hubble diagram for the $\Lambda$CDM model ($\Omega_{m}=0.3$, $\Omega_{\Lambda} = 0.7$) and H$_0 = 75.8$\,km\,s$^{-1}$\,Mpc$^{-1}$ (see Section \ref{txt:fiducial}). This Hubble diagram was built using the $i$-band magnitude, $(r-i)$ colour, and at 43\,d after the explosion. We present the number of SNe~II available at this epoch ($N_{\rm SNe}$) and the intrinsic dispersion ($\sigma_{\rm int}$). Note that the error bars do not include the intrinsic dispersion.}
\label{fig:HD_SCM}
\end{figure}

The method for determining H$_0$ can be divided into three steps. First, we need to calibrate the SN~II apparent magnitudes by deriving $\alpha$ and $\beta$ from Equation \ref{m_model}. To minimise the effect of peculiar-galaxy motions, we select SNe~II located in the Hubble flow, with $z_{\rm corr} > 0.01$ (89 objects). Figure \ref{fig:HD_SCM} shows the Hubble diagram for this sample. The uncertainties associated with the corrected magnitudes are
\begin{ceqn}
\begin{align}
\sigma^2_{m_{\rm corr}}=\sigma^2_{m}+\left(\frac{\alpha \sigma_{v_\mathrm{H\beta}}}{\mathrm{ln} 10 v_\mathrm{H\beta}}\right)^{2}+ (\beta \sigma_{(r-i)})^2+\sigma^2_{z},
\end{align}
\end{ceqn}
where $\sigma^2_{z}$ includes the redshift measurement uncertainties and a peculiar-velocity error of 250\,km\,s$^{-1}$. To the total uncertainty, a free parameter $\sigma_{\rm int}$ is added to take into account the unmodelled intrinsic SN~II scatter. A value of 0.27\,mag is derived, consistent with previous SCM research \citep{poznanski09,andrea10,dejaeger17b,dejaeger20}. The values of $\alpha$ and $\beta$ used to correct the SN~II apparent magnitudes are respectively $3.95^{+0.43}_{-0.42}$ and $1.07 \pm 0.28$ (see Section \ref{txt:fiducial}). If we assume that the colour-magnitude relation is due to extrinsic factors, the total-to-selective extinction ratio ($R_{V}$) can be obtained from $\beta$. We find a lower $R_{V} \approx 1$ than for SNe~Ia \citep{folatelli10}, but the low value could be due to intrinsic magnitude-colour not properly modelled. Recently, \citet{dejaeger18b} suggested that the majority of SN~II observed colour diversity is intrinsic and not produced by host-galaxy dust extinction. Note that these parameters depend on the sample chosen; in Section \ref{txt:samples}, we investigate their effects on the derived value of H$_0$. 

For the last two steps, we follow the work done by \citet{dhawan18} and adapt their Python programs\footnote{\url{https://github.com/sdhawan21/irh0/blob/master/full-analysis.ipynb}} to our SN~II sample. We derive the absolute magnitudes of all calibrators, $M^{\rm cal}_{i}$ ($\sigma_{M^{\rm cal}_{i}}$), using the Cepheid and TRGB distances from Table \ref{tab:calibrators} and by correcting their apparent magnitudes with the $\alpha$ and $\beta$ derived previously, 

\begin{ceqn}
\begin{align}
M^{\rm cal}_{i} = m^{\rm cal}_{i}+\alpha\, \mathrm{\log_{10}}\left(v_\mathrm{H\beta}\right) - \beta (r-i)-\mu_{\rm cal},\\
\sigma_{M^{\rm cal}_{i}} = \sigma^2_{m_{i}}+\left(\frac{\alpha}{\mathrm{ln} 10}\frac{\sigma_{v_\mathrm{H\beta}}}{v_\mathrm{H\beta}}\right)^2 + (\beta \sigma_{(r-i)})^2+\sigma^2_{\mu_{\rm cal}}+\sigma^2_{\rm int}.
\label{Mabs_cal}
\end{align}
\end{ceqn}

\begin{figure}
	\includegraphics[width=1.0\columnwidth]{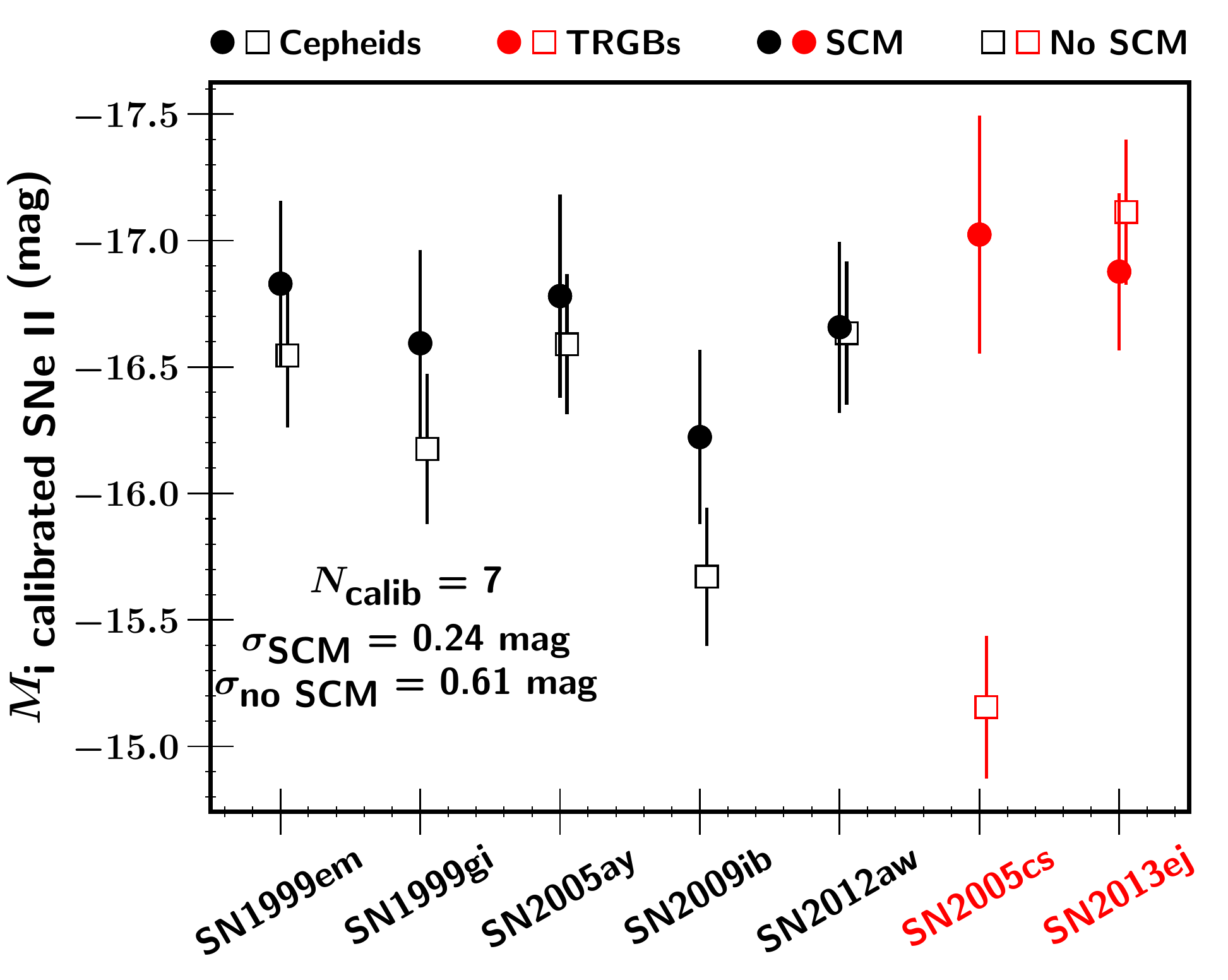}
\caption{Absolute $i$-band magnitude 43\,d after the explosion for the calibrators based on Cepheid distances (black) or the TRGB (red). The empty squares are the absolute magnitudes without applying the SCM, while the filled circles are with velocity and colour corrections. We also present the standard deviation with and without SCM.}
\label{fig:Mabs_cal}
\end{figure}

The absolute magnitudes for all seven calibrators are displayed in Figure \ref{fig:Mabs_cal}. Note that the uncertainties include the intrinsic scatter $\sigma_{\rm int}$. The calibrators have an average weighted absolute magnitude of $-16.69$\,mag with a dispersion of $\sigma_{\rm cal}= 0.24$\,mag. The dispersion is slightly larger than that obtained using SNe~Ia (0.16\,mag; \citealt{riess16,dhawan18}) but biased by the low-statistics number of calibrators (19 calibrators used for SNe~Ia). When the SCM is not applied, the dispersion increases to 0.61\,mag, demonstrating the utility of the SCM. SN~2005cs exhibits the largest difference with or without SCM because SN~2005cs is a low-luminosity SN~II \citep{pastorello06} and has small ejecta velocities, and therefore the largest $\alpha$\,log$_{10}$ (vel/<vel>) corrections.

It is important to note that on average, the absolute magnitude for the TRGBs is brighter ($N=2$, $-16.93 \pm 0.28$\,mag) than for the Cepheids ($N=5$, $-16.62 \pm 0.16$\,mag). As \citet{jang17a} showed that TRGB distances are in good agreement with the Cepheid distances derived by \citet{riess11,riess16}, the difference in absolute magnitude could be explained by a small-number statistics.

The third and last step consists of combining the calibrator and the Hubble-flow samples. The calibrator sample will constrain the absolute magnitude $M_i$, while the Hubble-flow sample is used to determined the intercept of the SN~II magnitude-redshift relation (zero-point). In practice,

\begin{ceqn}
\begin{align}
\mu=m_{i} - M_i = 5\,\log_{10}(d_{L}) + 25,
\label{eq:mu}
\end{align}
\end{ceqn}
where the luminosity distance ($d_{L}$) in Mpc is defined by its kinematic expression as
\begin{ceqn}
\begin{align}
d_{L} \approx \frac{cz}{H_{0}}\left( 1 + \frac{(1-q_{0})z}{2} + \frac{(1-q_{0}-3q_{0}^2+j_{0})z^{2}}{6} \right).
\label{eq:dl}
\end{align}
\end{ceqn}
\noindent
As defined by \citet{riess07}, $q_{0}$ is the present acceleration ($q_{0} = -0.55$) and $j_{0}$ is the prior deceleration ($j_{0} = 1$). Using Equations \ref{eq:mu} and \ref{eq:dl}, we can extract H$_0$ as
\begin{ceqn}
\begin{align}
\log_{10}~{H_0}= \frac{M_i + 5\,a_i + 25}{5},
\label{eq:H0}
\end{align}
\end{ceqn}
where $a_{i}$ is the intercept measured from the Hubble-flow sample and $M_{i}$ is the absolute SN~II $i$-band magnitude (at 43 d) derived using our calibrator sample. Following \citet{dhawan18}, to derive H$_0$, we use the Python package \textsc{EMCEE} developed by \citet{foreman13} and fit a joint model which combined the Hubble-flow and calibrator samples. The likelihood will evaluate how close the calibrators are to the mean absolute magnitude, and simultaneously how close the Hubble-flow SN~II absolute magnitudes are to the mean absolute magnitude given a value of H$_0$. In this model, $\alpha$, $\beta$, H$_0$, $M_i$, and $\sigma_{\rm int}$ are free parameters, and $a_i$ can be obtained using Equation \ref{eq:H0}. We run the MCMC simulation with 300 walkers and 2000 steps, and the priors are uniform for $\alpha$, $\beta$ $\neq$ 0, H$_0 > 0$, and $M_i < 0$, and scale-free for $0.0 < \sigma_{\rm int}$ with $p(\sigma_{\rm int})=1/\sigma_{\rm int}$.

\section{Results}\label{txt:results}

\subsection{Hubble constant}\label{txt:fiducial}

\begin{figure*}
\centering
\includegraphics[width=2.0\columnwidth]{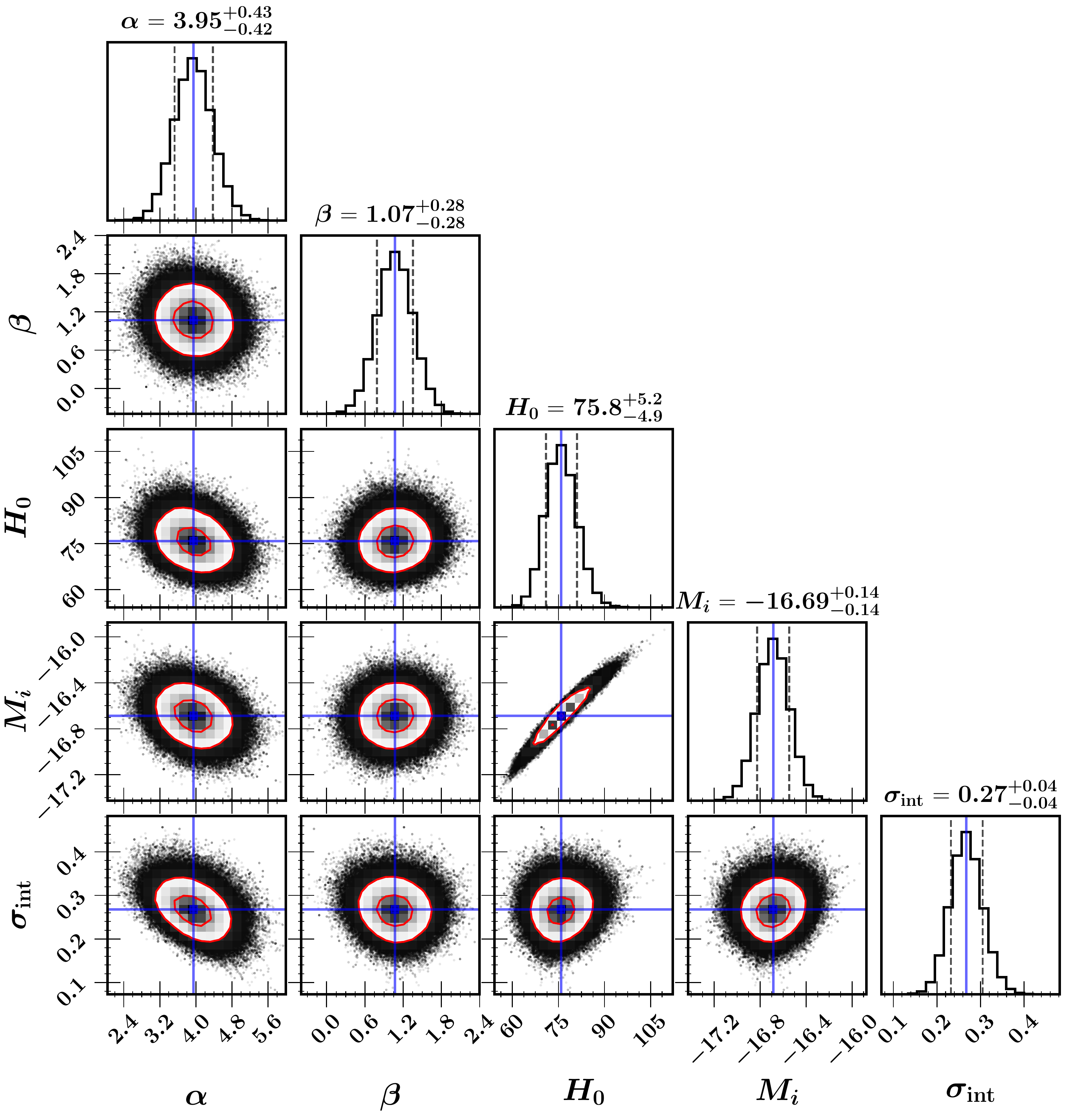}
\caption{Corner plot showing all of the one- and two-dimensional projections. Data points are shown as grey-scale points and red contours are given at 1$\sigma$ and 2$\sigma$ (which corresponds in two dimensions to the 39\% and 86\% of the volume). The five free parameters of our model are plotted: $\alpha$, $\beta$, H$_0$, $M_i$, and $\sigma_{\rm int}$. For each parameter, the median value and the 16th and 84th percentile difference are shown. To make this figure we used the corner-plot package (triangle.py v0.1.1. Zenodo. 10.5281/zenodo.11020).}
\label{fig:corner_plot_H0}
\end{figure*}

In Figure \ref{fig:corner_plot_H0}, the one- and two-dimensional projections of the five free parameters of our model ($\alpha$, $\beta$, H$_0$, $M_i$, and $\sigma_{\rm int}$) are shown. For the Hubble-flow SN~II sample, we use all SNe~II with $z_{\rm corr} > 0.01$ ($N=89$) and the seven calibrators described in Table \ref{tab:calibrators}. We obtain a median value of H$_0 = 75.8^{+5.2}_{-4.9}$\,km\,s$^{-1}$\,Mpc$^{-1}$, where the uncertainties are only statistical. With a $\sim 6.7$\% statistical uncertainty, this value is the most precise ever obtained using SNe~II.

Our result is consistent with the local H$_0$ determined from SNe~Ia ($74.03 \pm 1.42$\,km\,s$^{-1}$\,Mpc$^{-1}$; \citealt{riess19}) and shows a discrepancy of $\sim 1.7\sigma$ with the high-redshift value (H$_0 = 67.4 \pm 0.5$\,km\,s$^{-1}$\,Mpc$^{-1}$; \citealt{planck18}). The median absolute $i$-band magnitude 43\,d after the explosion of the calibrators is $M_i  = -16.69 \pm 0.14$\,mag while the intercept $-5\,a_i$ has a value of $-1.08 \pm 0.04$\,mag (derived using Eq. \ref{eq:H0}). Finally, an intrinsic scatter $\sigma_{\rm int} = 0.27 \pm 0.04$\,mag is obtained, consistent with previous SCM work where the community derived a value between 0.25 and 0.33 \citep{poznanski09,olivares10,andrea10,dejaeger17b,dejaeger20}.

\subsection{Systematic uncertainties}\label{txt:samples}

\begin{figure*}
\centering
\includegraphics[width=2.3\columnwidth]{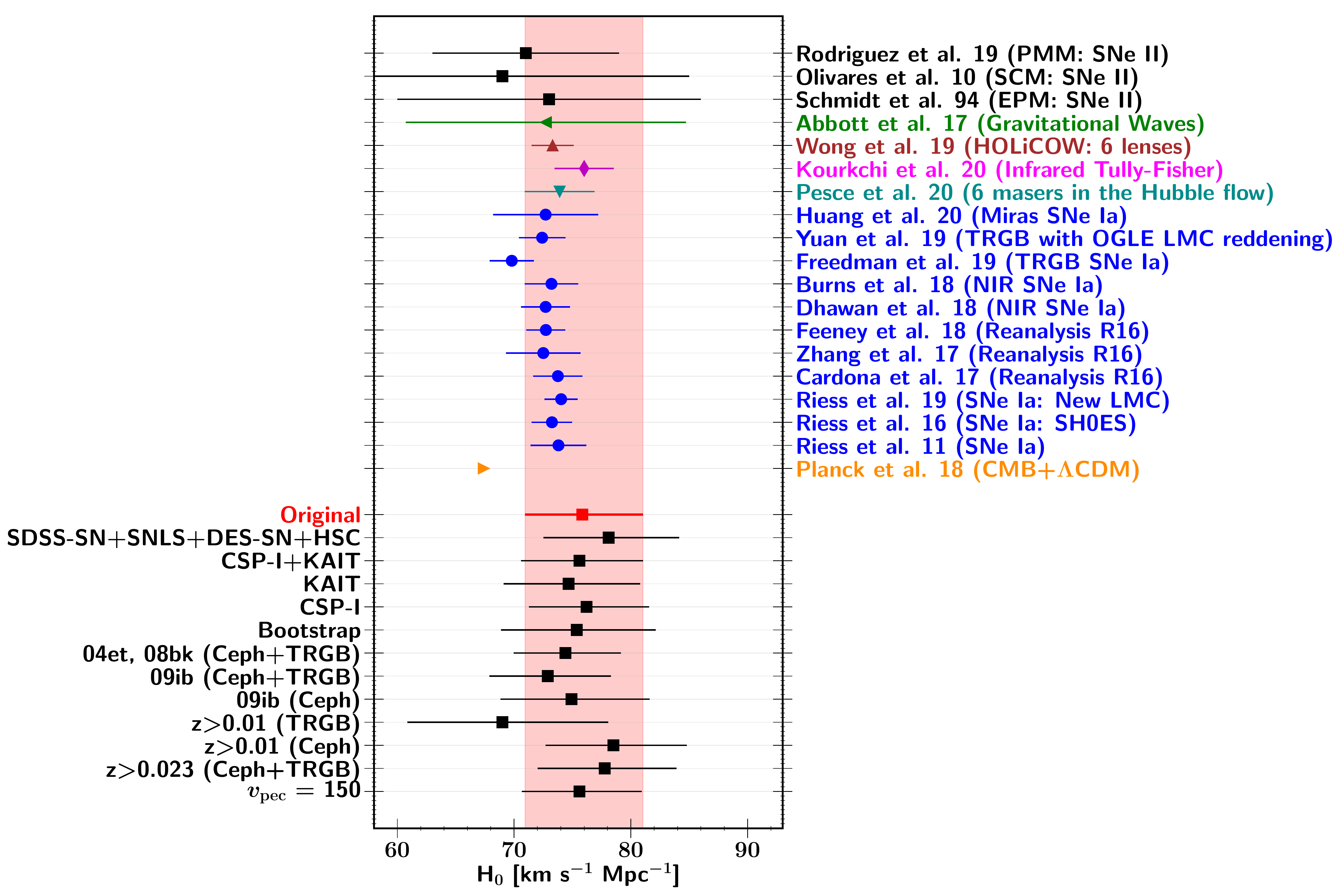}
\caption{Comparison of the H$_0$ value derived in this work using SNe~II with different cuts or samples (bottom black squares). The red filled region corresponds to the value obtained from our ``original'' sample ($z_{\rm corr} > 0.01$, $v_{\rm pec} = 250$\,km\,s$^{-1}$, and using our seven calibrators). We also added the local measurements from SNe~Ia (blue circles) using optical colours \citep{riess11,riess16,riess19}, the \citealt{riess16} data reanalysis \citep{cardona17,zhang17,feeney18}, using near-infrared (NIR) filters \citep{dhawan18,burns18}, with TRGB $+$ SNe~Ia \citep{freedman19}, TRGB $+$ SNe~Ia with OGLE LMC reddening \citep{yuan2019}, and with Mira variables $+$ SNe~Ia \citep{huang20}. Published values using SNe~II from \citet{schmidt94}, \citet{olivares10}, and \citet{rodriguez19a} are also shown (top black squares). H$_0$ values estimated using 6 masers in the Hubble flow \citep{pesce2020}, infrared Tully-Fisher relation \citep{kourkchi20}, the independent methods of quasar strong lensing \citep{wong2019}, and gravitational-wave sources \citep{abbott17} are also presented (dark cyan down-pointing triangle, magenta diamond, brown up-pointing triangle, and green left-pointing triangle, respectively). Finally, the high-$z$ value predicted by the CMB data $+$ $\Lambda$CDM from \citet{planck18} is represented by an orange right-pointing triangle.}
\label{fig:sys_H0}
\end{figure*}

In this section, we investigate the effect of our different cuts and calibrators on H$_0$; the results are summarised in Table \ref{tab:sys_H0} and Figure \ref{fig:sys_H0}.

First, if we change the peculiar velocity error to 150 \,km\,s$^{-1}$ (versus 250 \,km\,s$^{-1}$), H$_0$ slightly changes to $75.6^{+5.3}_{-4.9}$\,km\,s$^{-1}$\,Mpc$^{-1}$, only a difference of 0.3\% (0.2 \,km\,s$^{-1}$\,Mpc$^{-1}$). Then, if we select only the SNe~II with $z_{\rm corr} > 0.023$ \citep{riess11}, the number of SNe~II available for the Hubble diagram decreases to 47 and H$_0$ increases to $77.8^{+6.2}_{-5.7}$\,km\,s$^{-1}$\,Mpc$^{-1}$, a difference of 2.6\% (1.9\,km\,s$^{-1}$\,Mpc$^{-1}$). The difference can be explained by peculiar velocities that are not perfectly corrected. Increasing the sample size of SNe~II with $z_{\rm corr} > 0.023$ will reduce the systematic uncertainties caused by peculiar motions.

The largest difference in H$_0$ is seen when different calibrators are used. If only the Cepheids are selected as calibrators, H$_0$ increases to $78.5^{+6.3}_{-5.8}$\,km\,s$^{-1}$\,Mpc$^{-1}$, a difference of 3.2\% compared with the original sample. With the Cepheids, a slight difference of 0.5\% from the original sample is also seen for the absolute $i$-band magnitude: $-16.61 \pm 0.16$\,mag. As discussed in Section \ref{txt:H0_SNII} and Figure \ref{fig:Mabs_cal}, SN~II magnitudes calibrated with TRGBs are on average brighter than those calibrated using Cepheids. Therefore, if only the TRGBs are used as calibrators, the absolute $i$-band magnitude increases to $-16.89 \pm 0.27$\,mag, a difference of 0.20\,mag from the original sample. Although the dispersion decreases to 0.08\,mag (instead of 0.24\,mag), H$_0$ decreases to $69.0^{+9.1}_{-8.2}$\,km\,s$^{-1}$\,Mpc$^{-1}$, a difference of 9.0\%. However, the large difference is driven by small-number statistics. 

By removing the least luminous SN~II (SN~2009ib; see Fig. \ref{fig:Mabs_cal}) from the Cepheid sample, we derive a value smaller than that obtained with the original sample, $74.9^{+6.7}_{-6.1}$\,km\,s$^{-1}$\,Mpc$^{-1}$, a difference of only 0.9\,km\,s$^{-1}$\,Mpc$^{-1}$ ($\sim 1.2$\%). However, the absolute-magnitude dispersion of the calibrators decreases to 0.10\,mag (versus 0.21\,mag) but with only 4 SNe~II. We can also combine the Cepheid and TRGB samples after removing SN~2009ib. The calibrator absolute magnitude dispersion remains small (0.14\,mag) and the H$_{0}$ derived presents a difference of only 2.9 km s$^{-1}$ Mpc$^{-1}$ (3.8\%). As a test, we also add the two calibrators (SN~2004et and SN~2008bk) removed from our calibrator set (see Section \ref{txt:calibrator}). With the nine calibrators we obtain a value of $74.4^{+4.8}_{-4.5}$\,km\,s$^{-1}$\,Mpc$^{-1}$, which differs by 1.4\,km\,s$^{-1}$\,Mpc$^{-1}$ ($\sim 1.8$\%) 

\begin{figure}
\centering
\includegraphics[width=1.0\columnwidth]{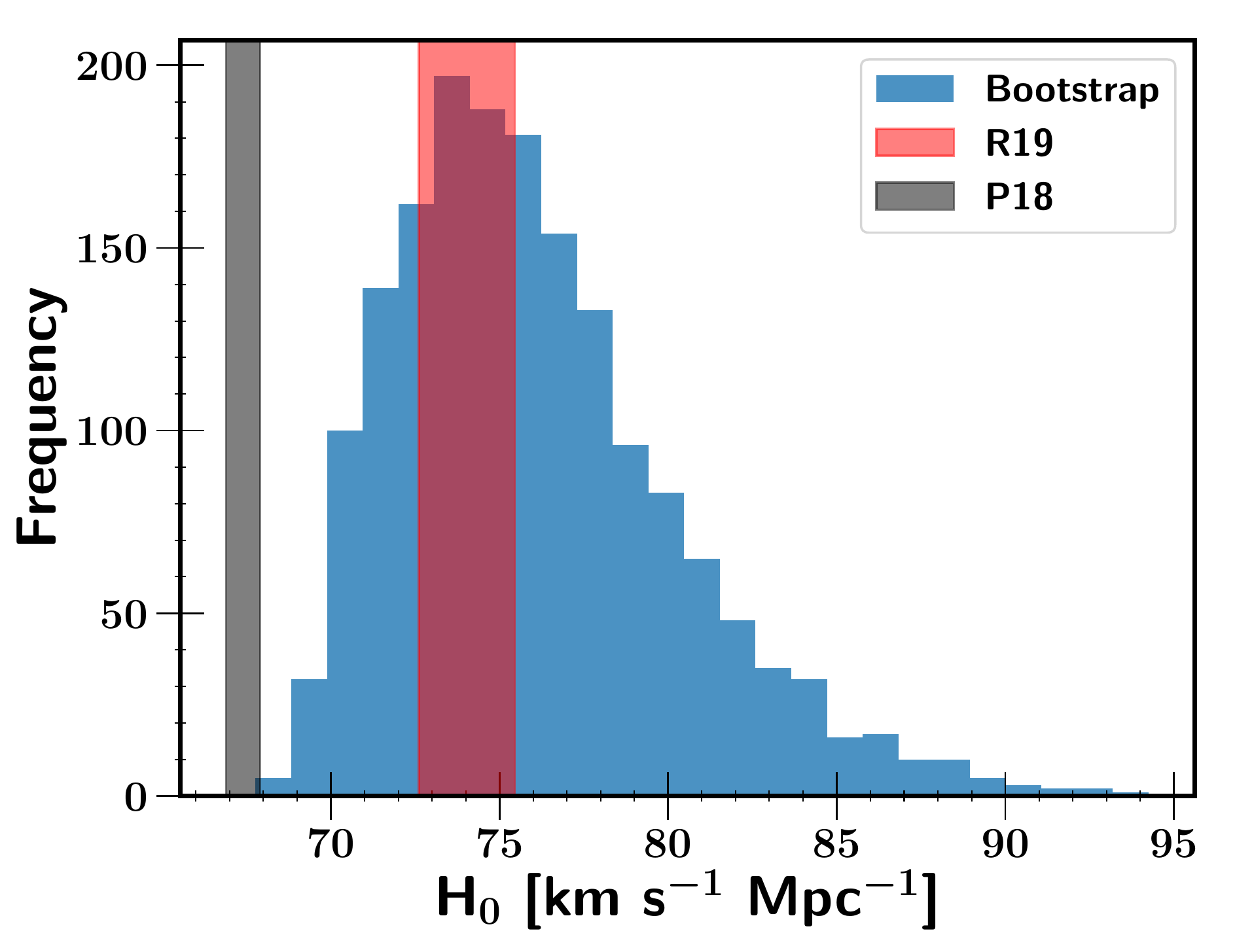}
\caption{Histogram of our bootstrap resampling of the set of calibrators, with replacement. In total, 1716 simulations were performed and 25 bins are used. An average value of $76.1 \pm 4.2$\,km\,s$^{-1}$\,Mpc$^{-1}$ and a median value of 75.4\,km\,s$^{-1}$\,Mpc$^{-1}$ are derived. The red and black filled regions correspond to the H$_0$ values obtained (respectively) by \citet{riess19} and \citet{planck18}. None of the 1716 H$_0$ values are smaller than 67.4 \,km\,s$^{-1}$\,Mpc$^{-1}$ \citep{planck18} and only 3.5\% have a discrepancy smaller than 1$\sigma$ (including only a systematic error of $2.8$\,km\,s$^{-1}$\,Mpc$^{-1}$).}
\label{fig:bootstrap}
\end{figure}

To get a more realistic idea of the calibrator effects on H$_0$, we also perform a bootstrap resampling of the set of calibrators, with replacement (see Figure \ref{fig:bootstrap}). To explore all the possibilities ($13!/7!6!$), we run 1716 simulations and get an average value of $76.1 \pm 4.2$\,km\,s$^{-1}$\,Mpc$^{-1}$. To compare with the original value, we take the median of these simulations, and as total uncertainty the standard deviation (4.2\,km\,s$^{-1}$\,Mpc$^{-1}$) added in quadrature to the mean of the errors obtained for each simulation ($+$5.3 and $-5.00$\,km\,s$^{-1}$\,Mpc$^{-1}$). We derive H$_0$ close to the original value: $75.4^{+6.8}_{-6.5}$\,km\,s$^{-1}$\,Mpc$^{-1}$, differing by only 0.4\,km\,s$^{-1}$\,Mpc$^{-1}$ ($\sim 0.5$\%). The distribution displayed in Figure \ref{fig:bootstrap} clearly shows that our H$_0$ value favours that obtained by \citet{riess19}. The peak of our distribution matches H$_0$ from the local measurements using SNe~Ia while the distribution almost does not overlap the {\it Planck}$+\Lambda$CDM value. The fact that our distribution extends to large H$_0$ values (85--95\,km\,s$^{-1}$\,Mpc$^{-1}$) is driven by SN~2009ib, which is the faintest calibrator. All the values larger than 85\,km\,s$^{-1}$\,Mpc$^{-1}$ were obtained when among the seven selected calibrators, SN~2009ib is used at least four times. For example, among the 1716 different possibilities, as we select seven SNe (with replacement) from our set of calibrators, for some combinations we used [SN 2009ib, SN 2009ib, SN 2009ib, SN 2009ib, X, Y, Z], where X, Y, and Z are SN~1999em, SN~1999gi, SN~2005ay, SN~2005cs, SN~2009ib, SN~2012aw, or SN~2013ej.

Finally, we can select different surveys to calibrate the SNe~II. If we select only CSP-I (37 SNe~II), H$_0$ is very consistent with the original sample value with a slight difference of 0.4\,km\,s$^{-1}$\,Mpc$^{-1}$ ($\sim 0.5$\%). Combining our two low-redshift samples (CSP-I and KAIT) or selecting only KAIT decreases the value to (respectively) $74.7^{+6.1}_{-5.6}$\,km\,s$^{-1}$\,Mpc$^{-1}$ and $75.6^{+5.4}_{-5.0}$\,km\,s$^{-1}$\,Mpc$^{-1}$; these are respective differences of 1.1\,km\,s$^{-1}$\,Mpc$^{-1}$ and 0.2\,km\,s$^{-1}$\,Mpc$^{-1}$ (1.5\% and 0.3\%). If we select only the ``high-$z$'' sample (SDSS-SN, SNLS, DES-SN, HSC), H$_0$ increases to $78.1^{+6.0}_{-5.6}$\,km\,s$^{-1}$\,Mpc$^{-1}$, a difference of 2.3\,km\,s$^{-1}$\,Mpc$^{-1}$ (3.0\%).

From the standard deviation of 13 analysis variants presented in Table \ref{tab:sys_H0}, we derive a systematic uncertainty of $\sim 2.5$\,km\,s$^{-1}$\,Mpc$^{-1}$ (i.e., 3.3\%). To this systematic uncertainty, we also add in quadrature an uncertainty of $\sim 1.2$\,km\,s$^{-1}$\,Mpc$^{-1}$ (1.5\% for the LMC in Table 6 of \citealt{riess19}) owing to the error in the anchor measurements (anchor distance, mean of period-luminosity relation in anchor, zero-points anchor-to-hosts, and Cepheid metallicity). The total systematic uncertainty is $\sim 2.8$\,km\,s$^{-1}$\,Mpc$^{-1}$ (i.e., 3.7\%). Thus, from our original sample, we obtain H$_0 = 75.8^{+5.2~{\rm(stat)}}_{-4.9~{\rm (stat)}} \pm 2.8~{\rm (sys)}$\,km\,s$^{-1}$\,Mpc$^{-1}$. Our H$_0$ value ($75.8 \pm 5.8$\,km\,s$^{-1}$\,Mpc$^{-1}$) derived using SNe~II is consistent (difference of $0.3\sigma$) with the local measurements from SNe~Ia \citep{riess19} but show a discrepancy of $\sim$ $1.4\sigma$ with the high-redshift results \citep{planck18}. Therefore, our value favours that obtained by \citet{riess19} (with a difference of only 1.8\,km\,s$^{-1}$\,Mpc$^{-1}$) rather than that estimated by \citet{planck18} (with a difference of 8.4\,km\,s$^{-1}$\,Mpc$^{-1}$). Note that the small calibrator set leads to large statistical uncertainty but leaves room for decreasing the total uncertainty. As a simple test, instead of selecting seven calibrators, we use each calibrator twice (14 calibrators in total). The final statistical uncertainty decreases from 5.1 (the average of 5.2 and 4.9) to 3.7\,km\,s$^{-1}$\,Mpc$^{-1}$ (25\%).

To summarise, in this work we only exchange SNe~II for SN~Ia to measure extragalactic distances; taking this study at face-value, there is no evidence that SNe~Ia are the source of the H$_0$ tension. The probability $P(H_0 \leq H_{0,Planck})$  that our H$_0$ measurement from the bootstrap distribution (moved toward the {\it Planck} measurement by the systematic error) is at least as low as 67.4 \,km\,s$^{-1}$\,Mpc$^{-1}$ \citep{planck18} is only 4.5\% (see Eq. 10 of \citealt{pesce2020}).

\begin{table*}
\leftskip=-1cm
\scriptsize
\caption{Free-parameter values for different sample choices.}
\begin{threeparttable}
\begin{tabular}{cccccccccccc}
\hline
Sample & Cali &$N_{\rm cali}$ &$\sigma_{\rm cali}$ &$N_{\rm SNe}$ &$\alpha$ &$\beta$ & H$_0$ &$M_i$ &$-5\,a_i$ & $\sigma_{\rm int}$ & $\Delta{{\rm H}_{0}}$\\
 &  & &(mag) & & & &(km\,s$^{-1}$\,Mpc$^{-1}$) &(mag) &(mag) & (mag) &\\
\hline
\hline
Original	&C$+$T 	&7	&0.24 &89 &3.95 $^{+0.43}_{-0.42}$ &1.07 $\pm$ 0.28 &75.8 $^{+5.2}_{-4.9}$ &$-$16.69 $\pm$ 0.14 &$-$1.08 $\pm$ 0.04 &0.27 $\pm$ 0.04 &$\cdots$\\
& & & & & & & & & & & \\ 
$v_{\rm pec}$= 150 km s$^{-1}$ &C$+$T 	&7 	&0.24 &89 &3.94 $^{+0.43}_{-0.42}$ &1.07 $^{+0.28}_{-0.27}$ &75.6 $^{+5.3}_{-4.9}$ &$-$16.69 $\pm$ 0.14 &$-$1.08 $\pm$ 0.04 &0.28 $\pm$ 0.04 &0.3\% \\
$z_{\rm corr}$ > 0.023 &C$+$T 	&7 	&0.29 &47 &4.07 $^{+0.65}_{-0.63}$ &0.41 $^{+0.48}_{-0.46}$ &77.8 $^{+6.2}_{-5.7}$ &$-$16.79 $\pm$ 0.16 &$-$1.24 $\pm$ 0.05 &0.28  $^{+0.05}_{-0.04}$ & 2.6\% \\
0.023 $>$ $z_{\rm corr}$ $>$ 0.15	&C$+$T 	&7 	&0.27 &40 &4.08 $^{+0.74}_{-0.72}$ &0.71 $^{+0.50}_{-0.49}$ &78.2 $^{+6.3}_{-5.8}$ &$-$16.74 $\pm$ 0.16 &$-$1.20 $\pm$ 0.06 &0.29  $\pm 0.05$ &3.2\% \\
$z_{\rm corr}$ > 0.01 &C 	&5 	&0.21 &89 &4.02 $^{+0.46}_{-0.45}$ &1.05 $^{+0.28}_{-0.29}$ &78.5 $^{+6.3}_{-5.8}$ &$-$16.61 $\pm$ 0.16 &$-$1.08 $\pm$ 0.04 &0.27  $\pm$ 0.04 &3.6\% \\
$z_{\rm corr}$ > 0.01 &T 	&2 	&0.08 &89 &3.96 $^{+0.44}_{-0.42}$ &1.00 $^{+0.30}_{-0.29}$ &69.0 $^{+9.1}_{-8.2}$ &$-$16.89 $\pm$ 0.27 &$-$1.09 $\pm$ 0.04 &0.27 $\pm$ 0.04 &9.0\% \\
$-$09ib	&C &4 &0.10 &89 &4.00 $^{+0.45}_{-0.44}$ &1.01 $^{+0.28}_{-0.29}$ &74.9 $^{+6.7}_{-6.1}$ &$-$16.71 $\pm$ 0.18 &$-$1.08 $\pm$ 0.04 &0.27 $\pm$ 0.04 &1.2\% \\
$-$09ib	&C$+$T &6 &0.14 &89 &3.94 $^{+0.42}_{-0.41}$ &1.03 $^{+0.28}_{-0.29}$ &72.9 $^{+5.4}_{-5.0}$ &$-$16.77 $\pm$ 0.15 &$-$1.08 $\pm$ 0.04 &0.27 $\pm$ 0.04 &3.8\% \\
$+$04et, 08bk	&C$+$T &9 &0.35 &89 &4.16 $^{+0.43}_{-0.42}$ &1.15 $^{+0.28}_{-0.29}$ &74.4 $^{+4.8}_{-4.5}$ &$-$16.73 $\pm$ 0.13 &$-$1.08 $\pm$ 0.04 &0.27 $\pm$ 0.03 &1.8\% \\
bootstrap &C$+$T &7 &0.18 &89 &3.96 $^{+0.45}_{-0.44}$ &1.04 $^{+0.28}_{-0.27}$ &75.4 $^{+6.8}_{-6.5}$ &$-$16.70 $\pm$ 0.18 &$-$1.08 $\pm $0.04 &0.27 $\pm$ 0.04 &0.5\% \\
CSP-I &C$+$T 	&7 &0.23 &37 &3.86 $^{+0.58}_{-0.56}$ &1.18 $^{+0.41}_{-0.40}$ &76.2 $^{+5.4}_{-4.9}$ &$-$16.62 $\pm$ 0.14 &$-$1.03 $\pm$ 0.05 &0.24 $\pm$ 0.04 &0.5\% \\
KAIT &C$+$T &7 &0.27 &19 &4.75 $^{+0.88}_{-0.93}$ &1.95 $^{+0.53}_{-0.56}$ &74.7 $^{+6.1}_{-5.6}$ &$-$16.59 $\pm$ 0.14 &$-$0.96 $\pm$ 0.08 &0.26 $\pm$ 0.07 &1.5\% \\
CSP-I$+$KAIT &C$+$T &7 &0.23 &56 &4.10 $^{+0.54}_{-0.50}$ &1.39 $^{+0.34}_{-0.35}$ &75.6 $^{+5.4}_{-5.0}$ &$-$16.62 $\pm$ 0.14 &$-$1.01 $\pm$ 0.05 &0.26 $\pm$ 0.05 &0.3\% \\
``high-$z$'' &C$+$T &7 &0.26 &33 &3.61 $^{+0.67}_{-0.64}$ &0.56 $^{+0.50}_{-0.49}$ &78.1$^{+6.0}_{-5.6}$ &$-$16.74 $\pm$ 0.15 &$-$1.21 $\pm$ 0.06 &0.27 $\pm$ 0.05 &3.0\% \\
%Double cal &C$+$T &14 &0.23 &89 &3.89 $^{+0.39}_{-0.39}$ &1.15 $^{+0.27}_{-0.27}$ &75.87 $^{+3.84}_{-3.61}$ &$-$16.68 $\pm$ 0.10 &$-$1.08 $\pm$ 0.04 &0.26 $\pm$ 0.04 &XXX\% \\
\hline
\hline
\end{tabular}
Effect of systematic errors on the best-fitting values using the SCM and different samples. Original line corresponds to the values obtained in Section \ref{txt:fiducial}. We try different cuts in redshift ($z_{\rm corr}$), surveys (e.g., only CSP-I, only KAIT, CSP-I$+$KAIT, high-$z$ SDSS$+$SNLS$+$DES$+$HSC), calibrators [Cepheids (C) and/or TRGBs (T)], and also remove or add some calibrators (e.g., $-$09ib for SN~2009ib; $+$04et, 08bk for SN~2004et, SN~2008bk). Bootstrap line corresponds to the average value obtained when performed a bootstrap resampling of the set of calibrators, with replacement using seven calibrators. For each parameter, the median value with the 16th and 84th percentile differences are given (uncertainties are only statistical). For bootstrap line, the median and the standard deviation added in quadrature to the mean of the uncertainties obtained for each parameter are written. The last column, $\Delta{{\rm H}_{0}}$, corresponds to the percentage difference from the original. 
\label{tab:sys_H0}
\end{threeparttable}
\end{table*}

\section{Conclusions}

In this work, we show that SNe~II have a role to play in the ``H$_0$ tension,'' as they can be used to obtain extragalactic distances and provide an independent measurement of H$_0$. From SNe~II and using only seven objects with Cepheid or TRGB independent host-galaxy distance measurements, we derive the most precise value of H$_0$ solely using SNe~II: H$_0 = 75.8^{+5.2}_{-4.9}$\,km\,s$^{-1}$\,Mpc$^{-1}$, where the uncertainties are only statistical.

We also investigate the effect of our different cuts and calibrators on H$_0$ and estimate a systematic error of $\sim 2.8$\,km\,s$^{-1}$\,Mpc$^{-1}$ ($\sim 3.7$\%). If we combine the systematic and the statistical errors, our value ($75.8 \pm 5.8$\,km\,s$^{-1}$\,Mpc$^{-1}$) is consistent with the local measurement \citep{riess19}, a difference of only 1.8\,km\,s$^{-1}$\,Mpc$^{-1}$. On the other hand, our H$_0$ value differs by $1.4\sigma$ from the high-redshift results \citep{planck18} -- a difference of 8.4\,km\,s$^{-1}$\,Mpc$^{-1}$. The probability that our H$_0$ measurement from the bootstrap distribution is at least as low as 67.4 \,km\,s$^{-1}$\,Mpc$^{-1}$ \citep{planck18} is only 4.5\%. Given that we only exchange SNe~II for SNe~Ia, this demonstrates that there is no evidence from our work that SNe~Ia are the source of the H$_0$ tension. However, it will be interesting to apply the inverse distance ladder method to SNe~II (CMB$+$BAO) and compare with the value of H$_0$ obtained in this work (distance-ladder method). Recent studies calibrating BAO$+$SNe~Ia from the Early Universe have obtained smaller Hubble constant values reinforcing the motion that the tension is an inconsistency between Late Universe geometry and Early Universe physics.

With the next generation of telescopes (e.g., the Vera C. Rubin Observatory), we will be able to increase the number of calibrators and SNe~II in the Hubble flow and thus improve the precision of H$_0$. For example, we showed that selecting only SNe~II with $z > 0.023$ in our present sample generates an H$_0$ variation of $\sim 3$\%.

\section*{Acknowledgements}

We thank the referee for their comments on the manuscript, which helped  improve it. Support for A.V.F.'s supernova research group at U.C. Berkeley has been provided by the NSF through grant AST-1211916, the TABASGO Foundation, Gary and Cynthia Bengier (T.d.J. is a Bengier Postdoctoral Fellow), Marc J. Staley (B.E.S. is a Marc J. Staley Graduate Fellow), the Christopher R. Redlich Fund, the Sylvia and Jim Katzman Foundation, and the Miller Institute for Basic Research in Science (U.C. Berkeley). L.G. was funded by the European Union's Horizon 2020 research and innovation programme under the Marie Sk\l{}odowska-Curie grant agreement No. 839090. This work has been partially supported by the Spanish grant PGC2018-095317-B-C21 within the European Funds for Regional Development (FEDER). The work of the CSP-I has been supported by the U.S. NSF under grants AST-0306969, AST-0607438, and AST-1008343.

KAIT and its ongoing operation were made possible by donations from Sun Microsystems, Inc., the Hewlett-Packard Company, AutoScope Corporation, Lick Observatory, the U.S. NSF, the University of California, the Sylvia \& Jim Katzman Foundation, and the TABASGO Foundation. Research at Lick Observatory is partially supported by a generous gift from Google. 
This research used the Savio computational cluster resource provided by the Berkeley Research Computing program at U.C. Berkeley (supported by the U.C. Berkeley Chancellor, Vice Chancellor for Research, and Chief Information Officer).

This paper is based in part on data collected at the Subaru Telescope and retrieved from the HSC data archive system, which is operated by the Subaru Telescope and Astronomy Data Center at the  National Astronomical Observatory of Japan (NAOJ). The Hyper Suprime-Cam (HSC) collaboration includes the astronomical communities of Japan and Taiwan, and Princeton University. The HSC instrumentation and software were developed by the NAOJ, the Kavli Institute for the Physics and Mathematics of the Universe (Kavli IPMU), the University of Tokyo, the High Energy Accelerator Research Organization (KEK), the Academia Sinica Institute for Astronomy and Astrophysics in Taiwan (ASIAA), and Princeton University. Funding was contributed by the FIRST program from the Japanese Cabinet Office, the Ministry of Education, Culture, Sports, Science and Technology (MEXT), the Japan Society for the Promotion of Science (JSPS), the Japan Science and Technology Agency (JST), the Toray Science Foundation, NAOJ, Kavli IPMU, KEK, ASIAA, and Princeton University.

The Pan-STARRS1 Surveys (PS1) were made possible through contributions of the Institute for Astronomy, the University of Hawaii, the Pan-STARRS Project Office, the Max-Planck Society and its participating institutes, the Max Planck Institute for Astronomy, Heidelberg and the Max Planck Institute for Extraterrestrial Physics, Garching, The Johns Hopkins University, Durham University, the University of Edinburgh, Queen's University Belfast, the Harvard-Smithsonian Center for Astrophysics, the Las Cumbres Observatory Global Telescope Network Incorporated, the National Central University of Taiwan, the Space Telescope Science Institute, the National Aeronautics and Space Administration (NASA) under grant No. NNX08AR22G issued through the Planetary Science Division of the NASA Science Mission Directorate, the U.S. NSF under grant AST-1238877, the University of Maryland, and Eotvos Lorand University (ELTE). This paper makes use of software developed for the Large Synoptic Survey Telescope. We thank the LSST Project for making their code available as free software at http://dm.lsst.org. 

Some of the data presented herein were obtained at the W. M. Keck Observatory, which is operated as a scientific partnership among the California Institute of Technology, the University of California, and NASA; the observatory was made possible by the generous financial support of the W. M. Keck Foundation. This work is based in part on data produced at the Canadian Astronomy Data Centre as part of the CFHT Legacy Survey, a collaborative project of the National Research Council of Canada and the French Centre National de la Recherche Scientifique. The work is also based on observations obtained at the Gemini Observatory, which is operated by the Association of Universities for Research in Astronomy, Inc., under a cooperative agreement with the U.S. NSF on behalf of the Gemini partnership: the U.S. NSF, the STFC (United Kingdom), the National Research Council (Canada), CONICYT (Chile), the Australian Research Council (Australia), CNPq (Brazil), and CONICET (Argentina). This research used observations from Gemini program numbers GN-2005A-Q-11, GN-2005B-Q-7, GN-2006A-Q-7, GS-2005A-Q-11, GS-2005B-Q-6, and GS-2008B-Q-56. This research has made use of the NASA/IPAC Extragalactic Database (NED), which is operated by the Jet Propulsion Laboratory, California Institute of Technology, under contract with NASA, and of data provided by the Central Bureau for Astronomical Telegrams.

Funding for the DES Projects has been provided by the U.S. Department of Energy, the U.S. NSF, the Ministry of Science and Education of Spain, the Science and Technology Facilities Council of the United Kingdom, the Higher Education Funding Council for England, the National Center for Supercomputing Applications at the University of Illinois at Urbana-Champaign, the Kavli Institute of Cosmological Physics at the University of Chicago, the Center for Cosmology and Astro-Particle Physics at the Ohio State University, the Mitchell Institute for Fundamental Physics and Astronomy at Texas A\&M University, Financiadora de Estudos e Projetos, Fundacao Carlos Chagas Filho de Amparo \'a Pesquisa do Estado do Rio de Janeiro, Conselho Nacional de Desenvolvimento Cient\'ifico e Tecnol\'ogico and the Minist\'erio da Ci\^encia, Tecnologia e Inovacao, the Deutsche Forschungsgemeinschaft and the Collaborating Institutions in the Dark Energy Survey.

The DES data management system is supported by the U.S. NSF under grant AST-1138766. The DES participants from Spanish institutions are partially supported by MINECO under grants AYA2012-39559, ESP2013-48274, FPA2013-47986, and Centro de Excelencia Severo Ochoa SEV-2012-0234. Research leading to these results has received funding from the European Research Council under the European Union's Seventh Framework Programme (FP7/2007-2013) including ERC grant agreements 240672, 291329, and 306478. This research uses resources of the National Energy Research Scientific Computing Center, a DOE Office of Science User Facility supported by the Office of Science of the U.S. Department of Energy under Contract No. DE-AC02-05CH11231.
 
\section*{Data Availability Statements}
The majority of the data have been already published and can be found in: \citet{poznanski09} (KAIT-P09), \citet{andrea10} (SDSS-SN), \citet{dejaeger17b} (HSC), \citet{dejaeger19} (KAIT-d19), and \citet{dejaeger20} (DES-SN).
CSP-I and SNLS data will be shared on reasonable request to the corresponding author.


%%%%%%%%%%%%%%%%%%%%%%%%%%%%%%%%%%%%%%%%%%%%%%%%%%
%%%%%%%%%%%%%%%%%%%%%%%%%%%%%%%%%%%%%%%%%%%%%%%%%%

% Don't change these lines
\bsp	% typesetting comment
\label{lastpage}
\end{document}